\renewcommand\a{\alpha}
\renewcommand\b{\beta}

\renewcommand\d{\delta}
\newcommand\e{\epsilon}

\newcommand\s{\sigma}

\newcommand\f{\phi}

\renewcommand\o{\omega}

\newcommand{\lan}{\langle}
\newcommand{\ran}{\rangle}

\newcommand{\non}{\nonumber\\}

\newcommand{\diracslash}[1]{#1\llap{/\kern2pt}}

\newcommand{\be}{\begin{equation}}
\newcommand{\ee}{\end{equation}}
\newcommand{\bea}{\begin{eqnarray}}
\newcommand{\eea}{\end{eqnarray}}
\newcommand{\ba}[1]{\begin{array}{#1}}
\newcommand{\ea}{\end{array}}
\newcommand{\bep}{\begin{pmatrix}}
\newcommand{\eep}{\end{pmatrix}}

\newcommand{\bt}{\begin{tabular}}
\newcommand{\et}{\end{tabular}}

\newcommand{\beas}{\begin{eqnarray*}}
\newcommand{\eeas}{\end{eqnarray*}}

\documentclass[preprint,prd,aps,floats,nofootinbib,floatfix]{revtex4}
\usepackage{graphicx}
\usepackage[intlimits]{amsmath}
\usepackage[utf8]{inputenc}
\usepackage[T1]{fontenc}
\usepackage{lineno}
\usepackage{textcomp}
\usepackage{subfig}
\usepackage{multirow}
\usepackage{xcolor} 
\usepackage{comment}
\usepackage[a4paper, total={160mm, 240mm}, top={0.5in}]{geometry}
\usepackage{tabularray}
\usepackage{ulem} 
\begin{document}
\title{Heavy Quarkonium-nuclear bound states within a generalized linear sigma model}
\author{Arpita Mondal}
\email{arpita.mondal@physics.iitd.ac.in, arpita.mondal1996@gmail.com
}
\affiliation{Department of Physics, Indian Institute of Technology, Delhi, Hauz Khas, New Delhi -- 110 016, India}

\author{Amruta Mishra}
\email{amruta@physics.iitd.ac.in}
\affiliation{Department of Physics, Indian Institute of Technology, Delhi, Hauz Khas, New Delhi -- 110 016, India} 
\begin{abstract}
We estimate the binding energies of charmonium ($J/\psi$, $\psi(2S)$, $\psi(1D)$, $\chi_{c0}$, $\chi_{c1}$, $\chi_{c2}$) and bottomonium ($\Upsilon(1S)$, $\Upsilon(2S)$, $\Upsilon_2(1D)$, $\chi_{b0}$, $\chi_{b1}$, $\chi_{b2}$) states bound in various nuclei (${\rm{^{4}He}}$, ${\rm{^{12}C}}$, ${\rm{^{16}O}}$, ${\rm{^{40}Ca}}$, ${\rm{^{90}Zr}}$, and ${\rm{^{208}Pb}}$) using the quarkonia-nuclei potentials obtained from their mass shifts in nuclear matter within the generalized linear sigma model. 
In the absence of light partons in heavy quarkonia, at the tree level, the medium modifications are driven by the gluon condensate, which is simulated within this model through a scalar dilaton field, $\chi$, by introducing broken scale invariance of QCD.
Our study shows that charmonium states bind more deeply with the atomic nuclei as compared to bottomonium states, providing a better probe for nuclear medium effects. Such bound states' investigations are particularly interesting for the upcoming J-PARC-E29, $\rm{\bar{P}ANDA}$@FAIR, and CEBAF@JLab experiments.
The mass shifts of the heavy quarkonium states in hot isospin asymmetric nuclear matter are investigated and are observed to receive an appreciable medium modification.
These medium effects are anticipated at FAIR@GSI, where such neutron-rich hot nuclear matter is expected to be produced.
\end{abstract}
\maketitle

\def\bfm#1{\mbox{\boldmath $#1$}}
\def\bfs#1{\mbox{\bf #1}}
\section{Introduction}
\label{intro}
The study of the strongly interacting matter is essential to understanding the complex dynamics of Quantum Chromodynamics (QCD) and its various phases that appear at different temperatures and baryon number densities.
For instance, at low temperatures, quarks and gluons get confined by the strong force, while at high temperatures, asymptotic freedom leads to a weakly coupled deconfined QCD medium of quarks and gluons. 
Experimental programs at the LHC (Large Hadron Collider) at CERN and the RHIC (Relativistic Heavy Ion Collider) at BNL have indicated the existence of a hot and dense phase of nuclear matter \cite{exp}.
Among the reliable signatures from experimental observations, heavy quarks (HQs), particularly charm and beauty, are considered as effective probes for investigating the properties of the nuclear matter produced during these collisions \cite{hqprobe}.
HQs are primarily created in the initial stages of the collisions at the time scale of $1/m_Q$ with $m_Q$ as the heavy quark mass.
They develop into physical resonance and due to the large mass, relative to the medium’s temperature, they do not thermalize fully.
Therefore, they witness the entire evolution of hot and dense matter, traversing plasma and nuclear matter before leaving the interacting system.
Thus, quarkonia at finite temperatures serve as valuable tools for probing the early stages of nuclear matter formation.

Several experiments have indicated that the properties of hadrons undergo medium modifications within the created strongly interacting matter.
For instance, the obtained dilepton spectra at the Super Proton Synchrotron (SPS) indicated the medium modifications of vector mesons \cite{ceres,helios,Brat,Cass, Mishra1,Mishra2}.
There have been extensive theoretical studies of these mesons within various frameworks including the potential models \cite{eichten_1, eichten_2, repko, Quarkonia_QGP_Mocsy_IJMPA28_2013_review}, QCD sum rule approach \cite{kimlee,klingl,hayashigaki,amarvjpsi_qsr,rajesh79,morita85,hatsuda}, coupled channel approach \cite{molina,tolos} , a chiral effective model \cite{prc98charmonium,AMC,DAM1,upsilon,AMCT},  and the quark meson coupling (QMC) model \cite{Tsushima, QMC_Krein, QMC1, QMC2}, indicating the ground as well as the excited state quarkonia properties are being modified within the created medium.
Our current investigation focuses on the in-medium effects on the masses of the S-wave ($\rm J/\psi$, $\rm \psi(2S)$ and $\rm \Upsilon(1S)$, $\rm \Upsilon(2S)$), P-wave ($\rm \chi_{c0}$, $\rm \chi_{c1}$, $\rm \chi_{c2}$ and $\rm \chi_{b0}$, $\rm \chi_{b1}$, $\rm \chi_{b2}$) and D-wave ($\rm \psi(1D)$ and $\rm \Upsilon_2(1D)$) sates of quarkonia (charmonia and bottomonia).
A significant fraction of quarkonia yield stems from the higher orbital quarkonia decays \cite{decay1,decay2}.
The masses of the heavy quarkonia, in the absence of light (anti)quark components, are expected to be modified due to the changes in the gluon condensates within the hot and dense nuclear medium.
Present study incorporates the effects of isospin asymmetry arising from the neutron-proton number imbalance along with the impacts of nonzero temperature and density.
As suggested in ref. \cite{leeko}, at the tree level, the in-medium masses are estimated through the leading order formula for the gluon condensate, assuming the distance between the heavy quark and antiquark is small compared to the scale of gluonic fluctuations. In our study, the gluon condensate is estimated using the generalized linear sigma model \cite{ellis, heide1992, Heide, fail}.
A standard linear sigma model is generalized by incorporating the scale invariance breaking through the logarithmic potential terms involving the scalar dilaton field $\chi$, the repulsive effects through $\omega$ meson, and the isospin asymmetry effects through $\rho$ meson in a chiral $SU(2)\times SU(2)$ Lagrangian.
Within this model, the scalar dilaton field $\chi$ simulates the gluon condensate through the trace matching condition.
Notably, such generalization leads to a reasonable description of the nuclear matter and finite nuclei properties \cite{Heide}.

Studying in-medium effects on the quarkonia masses can have significant observable consequences, such as particle yields, particle spectra, and collective flows, which are assessable in the upcoming HIC experiments using neutron-rich beams at ${\rm{\bar{P}ANDA}}$ and CBM experiments at FAIR (GSI) \cite{panda1, fair, cbm1, cbm2, panda2}, PHENIX Collaboration at RHIC at BNL \cite{phenix2}, LHCb Collaboration at CERN \cite{lhcb}, J-PARC \cite{jparc}, and NA61 Collaboration at CERN-SPS \cite{na61}.
These experiments are expected to unveil the in-medium properties of hadrons significantly.
Another potential approach to insight into the strongly interacting matter is by investigating the mesic-nuclei bound states.
Such states can occur when a meson is produced inside a nucleus with nearly zero recoil, receiving substantial attraction from the surrounding nuclear environment.
Experimental observation of such states could thus provide evidence of the mass reductions of the mesons within the nuclear medium.
This work includes the investigation of the probable bound states of the heavy quarkonia with ${\rm{^{4}He}}$, ${\rm{^{12}C}}$, ${\rm{^{16}O}}$, ${\rm{^{40}Ca}}$, ${\rm{^{90}Zr}}$, and ${\rm{^{208}Pb}}$ nuclei within the framework of generalized linear sigma model.
Currently, experimental programs at J-PARC-E29 \cite{e29i}, $\rm{\bar{P}ANDA}@$FAIR  \cite{fair1}, and CEBAF$@$JLab \cite{jlab1} are intended to produce low-momenta heavy quarkonia and open heavy flavor mesons in atomic nuclei.
However, experimentally it is challenging to meet such kinematic conditions.

In 1990, the binding of quarkonia to the nuclei was first proposed by S. J. Brodsky \textit{et. al.} \cite{brodsky}.
Till now, several studies have been conducted to investigate the possible existence of such exotic states.
In the ref. \cite{luke}, Luke \textit{et. al.} estimated the binding for $J/\psi$, in the limit of infinitely heavy quark mass, in the nuclear matter is around 10 MeV, whereas for $\Upsilon(1S)$ it is a few MeV and for the $\psi(2S)$ turned out to be an overestimation.
Later, in ref. \cite{leeko}, Lee and Ko calculated the binding energies for $J/\psi$, $\psi(2S)$, and $\psi(1D)$ in nuclear matter as around 8 MeV, 100 MeV, and 140 MeV, respectively, using the same mechanism with the finite charm quark mass and realistic charmonium bound-state.
Furthermore, the refs. \cite{qmc_C,qmc_B} estimated the binding energies for heavy quarkonia in the nuclear matter as well as finite nuclei through the excitation of the intermediate state hadrons, containing light (anti)quarks.
The results depend on the cut-off parameter used in the regularization process.
It varies around 18-24 MeV for $J/\psi$ and 16-22 MeV for $\Upsilon(1S)$ at saturation density ($\rho_0$). Therefore, the binding energies for these states in finite nuclei are almost comparable, precisely it occurs a little less for $\Upsilon(1S)$.
Using QCD sum rules \cite{kimlee, klingl, hayashigaki} the mass shift of $J/\psi$ in nuclear matter is estimated to be around -4 to -7 MeV.
A study by Mishra \textit{et. al.} \cite{amarvepja} showed that at zero temperature, the mass shifts are around -8.6(-8.4) MeV for $J/\psi$, -117(-114) MeV for $\psi(2S)$, and -155(-150) MeV for $\psi(1D)$ in symmetric(asymmetric) nuclear matter, using chiral SU(3) model.
Whereas, in ref.\cite{DAM1}, the obtained mass shifts of $\Upsilon(1S)$ and $\Upsilon(2S)$ are reported as -0.36(-0.34) MeV and -3.3(-3.1) MeV, respectively for $\eta=0$($\eta=0.5$).

The paper is organized as follows:  to begin with, in section \ref{model}, we briefly describe the generalized linear sigma model, explaining the nuclear matter (\ref{nm}) and the finite nuclei (\ref{fn}) scenario. 
Section \ref{med_mass} outlines the calculation of in-medium heavy quarkonia masses.
Next, in section \ref{qn_pot}, we address the condition where a quarkonium is inside a nucleus.
The model parameters, results, and discussions are presented in section \ref{result},  where we discuss the in-medium behavior of the fields, in-medium masses both in nuclear matter and finite nuclei, and the formation of the bound states of these mesons. 
Finally, in section \ref{summary}, we summarise our findings of the current study.
\section{GENERALIZED LINEAR SIGMA MODEL}
\label{model}
In this section, we describe a generalized linear sigma model \cite{ellis, heide1992, Heide, fail}, an effective Lagrangian approach based on the chiral $SU(2)\times SU(2)$ symmetry. Ideally, it is important to construct a Lagrangian both at the limit of exact symmetry and small symmetry breaking \cite{sreerup}. 
Therefore the linear sigma model is generalized to incorporate the QCD scale invariance breaking effect, through an effective potential in terms of the scalar dilaton field $\chi$ which simulates the scalar gluon condensate, $\Big<\frac{\alpha_s}{\pi}G_{\mu\nu}^aG^{a\mu\nu}\Big>$.
The Lagrangian includes an intermediate range attraction through scalar-isoscalar $\sigma$ meson and its chiral partner ${\vec\pi}(\equiv \pi^a,\;a=1,2,3)$ meson, and, a short-range repulsion through a vector-isoscalar $\omega_{\mu}$ meson and vector-isovector $\rho_{\mu}$ meson.
The Lagrangian density is given by \cite{Heide},
\bea
    \mathcal{L} = \mathcal{L}_0 - V_G + \mathcal{L'}.
    \label{c1}
\eea
The chiral symmetric, scale-invariant part of the Lagrangian density is,
\bea
    \mathcal{L}_0 =&\frac{1}{2}\partial_{\mu}\sigma \partial^{\mu}\sigma + \frac{1}{2}\partial_{\mu}{\pi}^a \partial^{\mu}{\pi}^a+ \frac{1}{2}\partial_{\mu}\chi\partial^{\mu}\chi - \frac{1}{4}\omega_{\mu\nu}\omega^{\mu\nu} + \frac{1}{2}\omega_{\mu}\omega^{\mu}[G_{\omega \sigma} (\sigma^2 + {\pi^a}^2)]\non &+ \bar{\psi}_N\Big(\gamma^{\mu}\Big(i\partial_{\mu} - g_{\omega}\omega_{\mu}\Big)- g_{\sigma}\Big(\sigma + i{\pi}^a{\tau}^a\gamma_5\Big) \Big)\psi_N; \quad
\rm{where,}\hspace{.5cm}
\psi_N = \left(\begin{array}{cc}
    \psi_p       \\
    \psi_n       
\end{array}\right),
 \label{c2}
\eea
$\omega_{\mu\nu} = \partial_{\mu}\omega_{\nu} - \partial_{\nu}\omega_{\mu}$ is the field strength tensor corresponds to the vector meson field $\omega_{\mu}$, and $\tau^a$ is the isospin operator. As suggested in ref. \cite{Heide}, in vacuum, $\omega$-meson mass takes the form $m_\omega =G_{\omega\sigma}^{1/2}\sigma_0$, with $\sigma_0$ as the vacuum expectation value of the $\sigma$ field, i.e., $\lan 0|\s|0\ran$.
The conventional Mexican hat potential of the linear sigma model fails to reproduce the fundamental nuclear ground state phenomenology.
Therefore, we have adopted the following potential $V_G (\chi, \sigma, \vec{\pi})$ which introduces the QCD scale symmetry-breaking effects through the inclusion of logarithmic terms of the scalar fields, 
  \bea
    V_G = B\chi^4 \left(ln\left(\frac{\chi}{\chi_0}\right) - \frac{1}{4}\right) - \frac{1}{2}B\delta \chi^4 ln\left({\frac{\sigma^2 + \vec{\pi}^2}{\sigma_0^2}}\right) + \frac{1}{2}B\delta\zeta^2\chi^2\Big(\sigma^2 +\vec{\pi}^2 - \frac{1}{2}\frac{\chi^2}{\zeta^2}\Big).
    \label{pot}
\eea
In Eq.(\ref{pot}), the dimensionless quantity $B$ is the bag parameter, and $\zeta$ is defined as, $\zeta = \chi_0/\sigma_0$.  The last term in Eq.(\ref{pot}) ensures, in the vacuum, the values of the fields will take $\chi = \chi_0$, $\sigma=\sigma_0$ and ${\pi}^a = 0$, representing the spontaneous breaking of chiral symmetry. The logarithmic terms are chosen to reproduce the trace anomaly. In the massless quark limit, the scalar gluon condensate breaks the scale symmetry, causing the non-zero trace of the energy-momentum tensor in QCD.
Within the model, the trace of the energy-momentum tensor becomes \cite{Heide, heide1992},
\bea
\langle \theta_{\mu}^{\mu} \rangle = 4 V_G(\Phi) - \sum \Phi \frac{\partial V_G}{\partial \Phi} = 4 \epsilon_{vac}\Big( \frac{\chi}{\chi_0} \Big)^4,
\label{c3}
\eea
where, $\Phi$ $\equiv$ \{$\chi,\sigma,\vec{\pi}$\} and the summation runs over all the scalar fields. The vacuum energy density $\epsilon_{vac}$ is estimated by the relation $\epsilon_{vac}$ = $-B\chi_0^4(1-\delta)/4$, which is determined by equating the trace of energy-momentum tensors obtained in the current framework and the chiral model \cite{amarvjpsi_qsr,DAM1}. With $N_c$ and the $N_f$ respectively the number of color and flavor quantum numbers of quarks, the parameter $\delta$ is given as $\delta=2N_f/(11N_c)$.
The dilaton field $\chi$ is associated with the scalar gluon condensate in the following way,
\bea
 \Big\langle \frac{\b(g)}{2g} G_{\mu\nu}^a G^{\mu\nu a} \Big\rangle = 4 \epsilon_{vac}\Big( \frac{\chi}{\chi_0} \Big)^4,
 \label{c5}
\eea
where $G^{a}_{\mu\nu}$ is the gluon field strength tensor and the relation is determined by equating the trace of the energy-momentum tensor calculated within the current model (see Eq.(\ref{c3})) with the one obtained from the general QCD Lagrangian, in the limit of zero current quark masses.
In relation (\ref{c5}), $\b(g)$ is the QCD renormalization group function, at the one-loop level, defined as,
\bea
\b(g) = -\frac{11N_c g^3}{48\pi^2}\Big(1-\frac{2N_f}{11N_c}\Big) + \mathcal{O}(g^5)~.
\label{c6}
\eea
It shows that the scalar gluon condensate is proportional to the fourth power of the scalar dilaton field, $\chi$. The vacuum mass of the nucleon as obtained from the Lagrangian (\ref{c2}) is $M = g \sigma_0$. According to the Goldberger-Treiman \cite {GT} relation, one can obtain the vacuum value of the sigma field as $\sigma_0 = f_{\pi}$ with $f_{\pi}$ is the pion decay constant, considering the axial vector coupling constant to be unity \cite{ellis}. Therefore, the vacuum value of the dilaton field becomes $\chi_0=\zeta \sigma_0\sim$ 148.8 MeV \cite{heide1992} and for $N_c = 3$, the relation between the gluon condensate and the dilaton field becomes,
\bea
\Big\langle \frac{\alpha_s}{\pi}G_{\mu\nu}^a G^{\mu\nu a}\Big\rangle = \frac{24}{33-2N_f} \Big( B(1-\delta) \chi^4 \Big)~;~~~ \rm{with}, ~\alpha_s = g^2/4\pi.
\label{c8}
\eea
The asymmetry between the number densities of proton and neutron in the created nuclear matter gives rise to the isospin asymmetric nuclear matter, which can be theoretically incorporated through the interaction of the nucleons with the vector-isovector field $\rho_{\mu}^a$. Here, $\mathcal{L'}$ is representing such interaction in the following way \cite{Heide},
\bea
    \mathcal{L'} = - \frac{1}{4}{\rho}_{\mu\nu}^a{\rho}^{\mu\nu,a} + \frac{1}{2}G_{\rho}\chi^2{\rho}_{\mu}^a{\rho}^{\mu,a} - g_\rho\bar{\psi}_N\gamma^{\mu}\frac{\tau^a}{2}\rho_{\mu}^a\psi_N~,
    \label{c9}
\eea
where, the field strength tensor corresponding to $\rho_{\mu}^a$ meson field is $\rho_{\mu\nu}^a = \partial_{\mu}\rho_{\nu}^a - \partial_{\nu}\rho_{\mu}^a $. In the vacuum the mass of the $\rho_{\mu}^a$ meson takes the form as $m_\rho =G_{\rho}^{1/2}\chi_0$.
By varying the total Lagrangian, the equations of motion of the fields $\sigma$, $\chi$, $\omega_{0}$, and $\rho_{0}^3$ can be obtained under the mean-field approximation (MFA), which neglects the Dirac sea contribution and thus the meson fields are treated as classical fields.
The equations and their respective source currents in different nuclear environments are discussed in the following subsections.
\subsection{For nuclear matter }
\label{nm}
The field equations within the hot, asymmetric nuclear matter are obtained from the Lagrangian density Eq.(\ref{c1}) under the MFA, where they attain ${\sigma} \rightarrow \lan{\sigma}\ran \equiv \sigma,\;{\chi} \rightarrow \lan{\chi}\ran \equiv \chi,\;{\pi^a} \rightarrow \lan{\pi^a}\ran \equiv 0, \; {\omega}_\mu \rightarrow\lan{\omega}_\mu\ran \equiv \d^{\mu 0}\,\omega_\mu = \o_0,\; {{\rho}}_{\mu}^a \to \lan{\rho}_{\mu}^a\ran \equiv \d^{\mu 0} \d^{a 3} \rho_{\mu}^a=\rho_0^3$, are given by,
\bea
    4B_0\phi^3(\ln{\phi} - \delta \ln{\nu})+ B_0 \delta\phi(\nu^2 - \phi^2)- m_{\rho}^2 \rho_0^2\phi &=& 0~,
    \label{c10}\\
 B_0\delta\left(\frac{\phi^4}{\nu} - \phi^2\nu\right) + m_{\omega}^2\omega_0^2\nu - M\sum_{i=p,n} \rho_i^s&=& 0~,
 \label{c11}\\
    m_{\omega}^2\nu^2 \omega_0 - g_\omega\sum_{i=p,n}\rho_i &=& 0~,
    \label{c12}\\
m_{\rho}^2\phi^2 \rho_0^3 - g_{\rho} \rho_3 &=& 0,
\label{c13}
\eea
with $B_0 = B\chi_0^4$ and the following source densities for the nucleons ($i$),
\bea
    \rho_i^s &=& \lan\bar{\psi}_i\psi_i\ran = \frac{\gamma_s}{(2\pi)^3}\int d^3k \frac{M_i^*}{\sqrt{k_i^2 + M_i^{*2}}} ~ [n_i(k) + \bar{n}_i(k)]~,
    \label{c14}\\
    \rho_i &=& \lan\psi_i^\dagger\psi_i\ran = \frac{\gamma_s}{(2\pi)^3}\int d^3k ~ [n_i(k) - \bar{n}_i(k)]~,
    \label{c15}\\
    \rho_3 &=& \frac{1}{2}[\lan{\psi}_p^{\dagger}\psi_p\ran - \lan{\psi}_n^{\dagger}\psi_n\ran] = \frac{1}{2}[\rho_p - \rho_n] = - {\eta} \rho_B,
    \label{c16}
\eea
where, $\gamma_s = 2$ is the spin degeneracy factor for protons and neutrons, $M_i^* = M_i \nu$ is the effective mass of the $i^{th}$ nucleon and $\eta = {(\rho_n - \rho_p)}/(2\rho_B)$ is the isospin asymmetry parameter. The temperature effects are incorporated through $n_i(k)$ and $\bar{n}_i(k)$, the fermion and antifermion distribution functions, respectively for the nucleons, $i$. At finite temperature \cite{song,fail} which forms as,
\bea
n_i(k) &=& \frac{1}{1+\exp[(E_i^*(k) - \mu_i^*)/T]}~,
\label{c16a}\\
\bar{n}_i(k) &=& \frac{1}{1+\exp[(E_i^*(k) + \mu_i^*)/T]}~,
\label{c16b}
\eea
with the effective nucleon energy $E_i^* = \sqrt{k_i^2+M_i^{*2}}$ and the effective chemical potential $\mu_i^*$, which is related to the chemical potential $ \mu_i$ of nucleons through the following relations \cite{mao},
\bea
\mu_p^* = \mu_p - g_\omega \omega_0 - \frac{1}{2} g_\rho \rho_0^3 ~,\\
\mu_n^* = \mu_n - g_\omega \omega_0 + \frac{1}{2} g_\rho \rho_0^3~,
\label{c17}
\eea
By solving the Eqs. (\ref{c10})-(\ref{c13}), the variations of the considered fields within the hot and dense medium can be obtained. Incorporating the in-medium scalar dilaton field, the medium modifications of the heavy quarkonia masses can be estimated, which shall be discussed in section \ref{med_mass}.
\subsection{For finite nucleus}
\label{fn}
In this subsection, we present the field equations for the finite nuclei, where the spatial dependence of the fields cannot be neglected as the system now has a finite spatial extent.
Thus the fields become $\s\rightarrow \s(r)$, $\nu\rightarrow \nu(r)$, $\chi\rightarrow \chi(r)$, $\phi\rightarrow \phi(r)$, $\o_0\rightarrow \o_0(r)$, $\rho_0^3\rightarrow \rho_0^3(r)$.
Moreover, while investigating the properties of the finite nuclei, it is necessary to include Coulomb interaction among the protons, which shall have less impact on the properties of the even nuclei than the odd nuclei, as reported in ref. \cite{me_fn_24}. Therefore, for the finite nuclei studies we consider the Lagrangian to be $\mathcal{L}+\mathcal{L}_C$, where $\mathcal{L}_C$ forms as,
\bea
\mathcal{L}_C = -\frac{1}{4}A_{\mu\nu}A^{\mu\nu} - \bar{\psi}_N\gamma^\mu\Big(\frac{e}{2}(1+\tau^a)A_{\mu} \Big)\psi_N,\;\;\;\;\; {\rm{with,}}\;\; A_{\mu\nu} = \partial_{\mu}A_{\nu} - \partial_{\nu}A_{\mu},
\eea
where, under MFA, the Coulomb field becomes $A_\mu \rightarrow\lan{A}_\mu\ran \equiv \d^{\mu 0}\, A_\mu=A_0$.
The following equations, obtained from the considered Lagrangian, represent the time-independent behaviors of the mean fields within static, spherically symmetric nuclei,
\bea
    \chi_0^2\mathcal{D}\f &=& 4B_0\phi^3(\ln{\phi} - \delta \ln{\nu})+ B_0 \delta\phi(\nu^2 - \phi^2)- m_{\rho}^2 (\rho_0^3)^2\phi,
    \label{f10}\\
\s_0^2\mathcal{D}\nu &=& M\sum_{i=p,n}\rho_i^s - B_0\delta\left(\frac{\phi^4}{\nu} - \phi^2\nu\right) - m_{\omega}^2\omega_0^2\nu,
 \label{f11}\\
   \mathcal{D}\o_0 &=& -g_{\omega}\sum_{i=p,n}\rho_i +  m_{\omega}^2\nu^2 \omega_0,
    \label{f12}\\
\mathcal{D}\rho_0^3 &=& - g_{\rho} \rho_3 + m_{\rho}^2\phi^2 \rho_0^3.
\label{f13}\\
\mathcal{D}A_0 &=& -e\rho_p,
\label{f14}
\eea
with $\mathcal{D} \equiv \frac{d^2}{dr^2} + \frac{2}{r}\frac{d}{dr}$ and the corresponding proton and neutron densities are defined as,
\bea
\label{ps_den}
\rho^s_p(r) 
&=&  \sum_{p}^Z \bar{\varphi}_p(r)\varphi_p(r)
= \sum_{p}^Z \frac{(2j_p+1)}{4\pi r^2}(|G_p(r)|^2 - |F_p(r)|^2),\\
\label{ns_den}
\rho^s_n(r) 
&=&  \sum_{n}^N \bar{\varphi}_n(r)\varphi_n(r)
= \sum_{n}^N \frac{(2j_n+1)}{4\pi r^2}(|G_n(r)|^2 - |F_n(r)|^2),\\ \label{p_den}
\rho_p(r) 
 &=&  \sum_{p}^Z {\varphi}^{\dagger}_p(r) \frac{(1+\tau^3_p)}{2}\varphi_p(r)= \sum_{p}^Z \frac{(2j_p+1)}{4\pi r^2}(|G_p(r)|^2 + |F_p(r)|^2),\\ \label{n_den}
\rho_n(r) 
 &=&  \sum_{n}^N {\varphi}^{\dagger}_n(r) \frac{(1-\tau^3_n)}{2}\varphi_n(r)
= \sum_{n}^N \frac{(2j_p+1)}{4\pi r^2}(|G_n(r)|^2 + |F_n(r)|^2),
\eea
where, $Z$ and $N$ are the proton number and neutron number of the nucleus, respectively and $\tau^3_N$ is the isospin operator. The summation is performed only over the occupied orbits in the Fermi sea, since, the densities are evaluated in the no-sea approximation, by considering only the following positive energy Dirac spinors,
\begin{align}
\label{diracspinor}
\varphi_{\a}^\kappa ({{r}}) = \begin{pmatrix}
    i\frac{G_\a^\kappa (r)}{r} Y^l_{jm} (\theta,\phi)\\
    -\frac{F_\a^\kappa (r)}{r} Y^{\tilde{l}}_{jm} (\theta,\phi)
\end{pmatrix}  \chi_{t_\a} (t), 
\end{align}
where, $iG_\a(r)/r$ and $-F_\a(r)/r$ are respectively the radial part of the upper and lower components, characterized by the single particle quantum numbers $\a$, orbital momentum quantum numbers $l$, angular momentum quantum numbers $j,m$ and $\kappa$ ($=(-1)^{j+l+1/2}(j+1/2)$) and the isospin projection $t_\a = \tau_\a^3/2$, with $\chi_{t_\a}$ being two component Pauli spinor and $Y^{l,\tilde{l}}_{jm}$ are the spin spherical harmonics where $\tilde{l} = l+ (-1)^{j+l-1/2}$, can be evaluated from the following Dirac equation of the nucleons,
\bea
\label{diraceq}
\Big(\frac{d}{dr} + \frac{\kappa}{r}\Big) G_\a(r) - \Big[\e_\a - \Big(V_\o(r) + \frac{\tau^3_\a}{2}V_\rho(r) + \frac{1+\tau^3_\a}{2}V_C(r)\Big) + V_\s(r)\Big]F_\a(r) = 0, \non
\Big(\frac{d}{dr} - \frac{\kappa}{r}\Big) F_\a(r) + \Big[\e_\a - \Big(V_\o(r) + \frac{\tau^3_\a}{2}V_\rho(r) + \frac{1+\tau^3_\a}{2}V_C(r)\Big) - V_\s(r)\Big]G_\a(r) = 0,
\eea
where the nucleons are moving independently in the mean field potentials $V_\sigma(r) = g_{\s} \s(r)$, $V_{\omega}(r) = g_\omega \omega_0(r)$, $V_{\rho}(r) = g_{\rho}\rho_{0}^3(r)$, $V_C(r) = eA_0(r)$ and $\e_\a$ is the energy, related to the single-particle energy ($E_\a$) through the expression $\e_\a = E_\a -M $.
Following the method stated in ref. \cite{me_fn_24}, the behaviors of the fields inside an atomic nucleus can be estimated by solving the Eqs. (\ref{f10})-(\ref{diraceq}) self-consistently.
\section {IN-MEDIUM MASSES OF THE HEAVY QUARKONIA}
\label{med_mass}

The medium modifications of the masses of the heavy quarkonium states are inherited solely from the in-medium scalar gluon condensate due to the absence of light quarks and antiquarks in its structure.
In ref. \cite{pes1}, M. E. Peskin and in ref. \cite{pes2}, G. Bhanot and M. E. Peskin described the interaction between a nucleon and a heavy quarkonium state in the nuclear medium using perturbative quantum chromodynamics (pQCD). 
The calculations are based on the leading order (LO) operator product expansion (OPE) of the correlation function between two heavy quark currents.
In the present work, for heavy quark systems, the OPE is limited up to the lowest dimensional (dimension-4) operator, the scalar gluon condensate \Big($\big\lan(\a_s/\pi) G_{\mu\nu}^a G^{\mu\nu a}\big\ran$\Big).
Using pQCD, the LO effect of the in-medium gluon condensates on the masses of the quarkonium states in the nuclear medium, for the large heavy quark mass limit is presented by S. H. Lee and C. M. Ko \cite{leeko}.
With the assumptions, (i) the wave function of the quarkonium is Gaussian, (ii) in the momentum space, which is normalized as $\int{\frac{d^3k}{(2\pi)^3}|\psi(\bf k)|^2} = 1$, and (iii) the distance between the heavy quark and antiquark, which are being bound by a Coulomb potential, is small as compared to the scale of gluonic fluctuations, the form of the mass shift is given by,
\bea
    \Delta m = \frac{1}{18}\int d{k}^2 \Big\langle \Big\vert \frac{\partial \psi (\bf k)}{\partial {\bf k}} 
\Big\vert^{2} \Big\rangle \frac{k}{\frac{{k}^2}{m_Q} + \epsilon}\Big(\Big<\frac{\alpha_s}{\pi} G_{\mu\nu}^a G^{\mu\nu a}\Big> - \left<\frac{\alpha_s}{\pi} G_{\mu\nu}^a G^{\mu\nu a}\Big>_0\right),
\label{Q1}
\eea
where,
\bea
\Big\langle \Big\vert \frac{\partial \psi (\bf k)}{\partial {\bf k}} 
\Big\vert^{2} \Big\rangle
=\frac {1}{4\pi}\int 
\Big\vert \frac{\partial \psi (\bf k)}{\partial {\bf k}} \Big\vert^{2}
d\Omega, \;\;\;\; {\rm{with,}}\;\; k = |\bf k|
\label{Q2}
\eea
and, $m_Q$ is the heavy quark ($Q$) mass, taken as 1.95 (5.36) GeV for charm (beauty) quark to reproduce the energy splitting between the $1S$ and $2S$ states in the vacuum and the binding energy of each respective states properly. 
The vacuum mass of the corresponding quarkonium state is denoted by $m$ and the binding energy of the heavy quark-antiquark bound state is, $\epsilon=2m_Q-m$. 
The wave functions of the different quarkonium states are determined by treating them as a quark-antiquark bound state under a harmonic oscillator potential \cite{leeko}, can be obtained by solving the following Schr\"odinger equation \cite{arfken},
\bea
\nabla^2\psi + \frac{2M_Q}{\hbar^2}\Big(E - \frac{1}{2}M_Q\omega^2 r^2\Big)\psi = 0,
\label{W1}
\eea
where $M_Q = m_Q/2$ is the reduced mass of the heavy quarkonium system and the obtained wave function is \cite{FLS},
\bea
    \psi_{N,l}(r,\theta,\phi) = A\ Y_l^m (\theta, \phi) (\beta^2 r^2)^\frac{l}{2} e^{-\frac{1}{2}\beta^2r^2} L^{(l+\frac{1}{2})}_{N-1}(\beta^2r^2),
    \label{W2}
\eea
with, $\vec{r} = \vec{r}_Q - \vec{r}_{\bar{Q}}$, is the relative radial coordinate and A is the normalization constant.
Here, the parameter $\beta$ characterizes the strength of the harmonic oscillator potential for the respective quarkonium states, expressed as $\beta^2 = M_Q\omega/{\hbar}$ \cite{FLS}, can be fixed by the analytic expression of r.m.s. radii of the respective quarkonium states, i.e., $<r^2>^\frac{1}{2} = (\int d^3r \psi^* r^2 \psi)^\frac{1}{2}$, and $L_a^b(x)$ is the associated Laguerre Polynomial \cite{beta1}.
The derivative term of the wave function in Eq.(\ref{Q1}) represents the measurement of the color dipole size of the corresponding state.
Within the model, after incorporating the relation (\ref{c8}) for $N_c = 3$ and $N_f = 2$, the masses of the heavy quarkonia in the nuclear medium receive the medium modification through the dilaton field $\chi$ and the form of the mass shift is revised to,
\bea
  \Delta m = \frac{4}{87} B(1-\delta)\int d{\bf k}^2 \Big\langle\Big|\frac{\partial \psi(\bf k)}{\partial {\bf k}}\Big|^2\Big\rangle \frac{\bf k}{\frac{{\bf k}^2}{m_Q} + \epsilon}\Big( \chi^4 - \chi_0^4 \Big),
  \label{Q4}
\eea
which states that the mass shift is proportional to the difference in the values of the fourth power of the scalar dilaton field $\chi$ in medium and vacuum.

Notably, when the heavy quarkonium states are captured inside the nuclei, after its slow production, the mass variation inside the nuclei becomes implicitly position dependent through the $\chi(r)$ field, where $\vec{r}$ is the position from the center of the nucleus with $r =|\vec{r}|$.
This can be approached in two ways. 
First, once the behavior of the dilaton field in the nucleus is determined, the quarkonia masses within the nucleus can be directly estimated from Eq. (\ref{Q4}).
Alternatively, after obtaining the mass behavior in infinite nuclear matter, the local density approximation (LDA) can be applied by replacing $\rho$ by $\rho(r)$, the local density at each point in the nucleus, calculated from finite nucleus calculation within the current study, the behavior of the mass shifts within a finite nucleus can be achieved through the relation, $\Delta m(r)\equiv\Delta m(\rho(r))$.

\section{HEAVY QUARKONIA-NUCLEI POTENTIALS}
\label{qn_pot}
This section lays out the potentials experienced by the heavy quarkonia while they are produced with low momentum inside the nucleus. 
Since the quarkonium consists of $Q-\bar{Q}$, i.e., the same flavor of quark and antiquark, they feel equal and opposite Lorentz vector potentials.
Therefore the only potential experienced by the quarkonium is $\Delta m(r)$, the scalar potential.
With a careful inspection of the effective potential or the in-medium mass shifts, the existence of the quarkonium bound states with the considered nucleus can be estimated naively. 
Moreover, a negative binding energy and a well-behaved eigenstate within the nucleus for the given potentials are the signatures of the bound states.
To obtain the binding energies of quarkonia in the nucleus and the corresponding eigenfunctions, we solve the following relativistic Klein-Gordon equation for the mesons using the scalar nuclear potential for various nuclei \cite{qmc_B},
\begin{align}
\label{final}
 \big[\nabla^2 + \epsilon^2 - m^{\star 2 }(r)\big] \Phi(r) = 0, \;\;\; {\rm{where,}}\;\;\nabla^2 = \frac{d^2}{dr^2} - \frac{l(l+1)}{r^2}~ ,
\end{align}
$m^*(r) = m+\Delta m(r)$, $\epsilon$ is the binding energy of the state, which is related to the single-particle energies as $E = (\epsilon - m)$ and $\Phi(r)$ is the corresponding state wave function.
As the width of these quarkonia states in free space is very small ($\rm{\sim keV}$), the strong interaction width of these mesons is considered to be zero in the effective potential, which keeps the Eq. (\ref{final}) real. 
\section{Results and Discussions}
\label{result}
The mass shifts of the charmonium
states $J/\psi\rm{(1^3S_1)}$, $\psi(2S)\rm{(2^3S_1)}$, $\psi(1D)\rm{(1^3D_1)}$, $\chi_{c0}\rm{(1^3P_0)}$, $\chi_{c1}\rm{ (1^3P_1)}$ and $\chi_{c2}\rm{(1^3P_2)}$ with their vacuum masses (in MeV), 3096.9, 3686.1, 3773.7, 3414.7, 3510.7 and 3556.2, respectively and the bottomonium states $\Upsilon(1S)\rm{(1^3S_1)}$, $\Upsilon(2S) \rm{(2^3S_1)}$, $\Upsilon_2(1D)\rm{ (1^3D_2)}$, $\chi_{b0}\rm{(1^3P_0)}$, $\chi_{b1}\rm{(1^3P_1)}$ and $\chi_{b2}\rm{(1^3P_2)}$ with the free space masses (in MeV), 9460.3, 10023.3, 10163.7, 9859.4, 9892.8 and 9912.2, respectively \cite{pdg}, are studied in the hot and dense symmetric as well as asymmetric nuclear matter within the generalized linear sigma model.
The mass shifts of the quarkonia are obtained through the modifications of the scalar dilaton field ($\chi$) within the nuclear medium through Eq. (\ref{Q4}).
In the present work, we also explore the probable formation of the bound states of these quarkonia with ${\rm{^{4}He}}$, ${\rm{^{12}C}}$, ${\rm{^{16}O}}$, ${\rm{^{40}Ca}}$, ${\rm{^{90}Zr}}$, and ${\rm{^{208}Pb}}$ nuclei, using the Klein-Gordon equation for the obtained meson–nucleus potentials.
Using the parameters, discussed in subsection \ref{Param}, the in-medium behavior of the scalar dilaton field and the scalar isoscalar field as a function of nuclear matter density in the symmetric ($\eta=0$) and asymmetric ($\eta=0.5$) nuclear matter, at various temperatures, by solving the set of coupled equations of motion in Eqs. (\ref{c10}) to (\ref{c13}) are studied in subsection \ref{A}. 
Finally, in subsection \ref{B}, we discuss the mass shifts of the charmonium and bottomonium states in the hot and dense nuclear matter.
In the same subsection, the behaviors of these mass shifts in the finite nuclear environment are described, followed by their probable formation of bound states with the nuclei explored in the next subsection \ref{bound}.
\subsection{Parameters} \label{Param}
Before plunging into the results and the corresponding analyses, we first state the parameters chosen in the present work. In section \ref{model}, all the parameters are introduced briefly at an introductory level. The parameter $C_{\omega}^2 = g_{\omega}^2 M^2/m_{\omega}^2$ can be chosen to fit the nuclear matter saturation properties and $C_{\rho}^2 = g_\rho^2 M^2/m_\rho^2$ is fixed by the bulk symmetry energy of 35 MeV \cite{Heide}, with $m_\o$, $m_\rho$ and $M$ as 783 MeV, 770 MeV, and 939 MeV, respectively. From the one-loop approximation of the QCD $\beta$ function, the suggested $\delta $ parameter is 4/33, considering color quantum number $N_c = 3$ and flavor quantum number $N_f = 2$. The parameters \cite{Heide, fail} are presented in Table- \ref{table1}.

\begin{table}[h!]
\vspace{0.5cm}
\centering
\begin{tabular}{c  c  c  c  c  c }
\hline
 $\delta$ & \qquad $\big|\epsilon_{vac}\big|^{\frac{1}{4}}$(MeV) & \qquad $C_{\omega}^2$ & \qquad $C_\rho^2$ & \qquad $\sigma_0$(MeV) & \qquad $\chi_0$(MeV)\\
\hline
$\frac{4}{33}$ & \qquad 269 & \qquad 51.3 & \qquad 132 & \qquad93 & \qquad 148.8\\
\hline
\end{tabular}
\caption{ \raggedright{Values of the different parameters.}}
\label{table1}
\end{table}
\subsection{Behaviour of the scalar fields in nuclear matter}
\label{A}
\begin{figure}[th]
\centerline{\includegraphics[width=14cm]{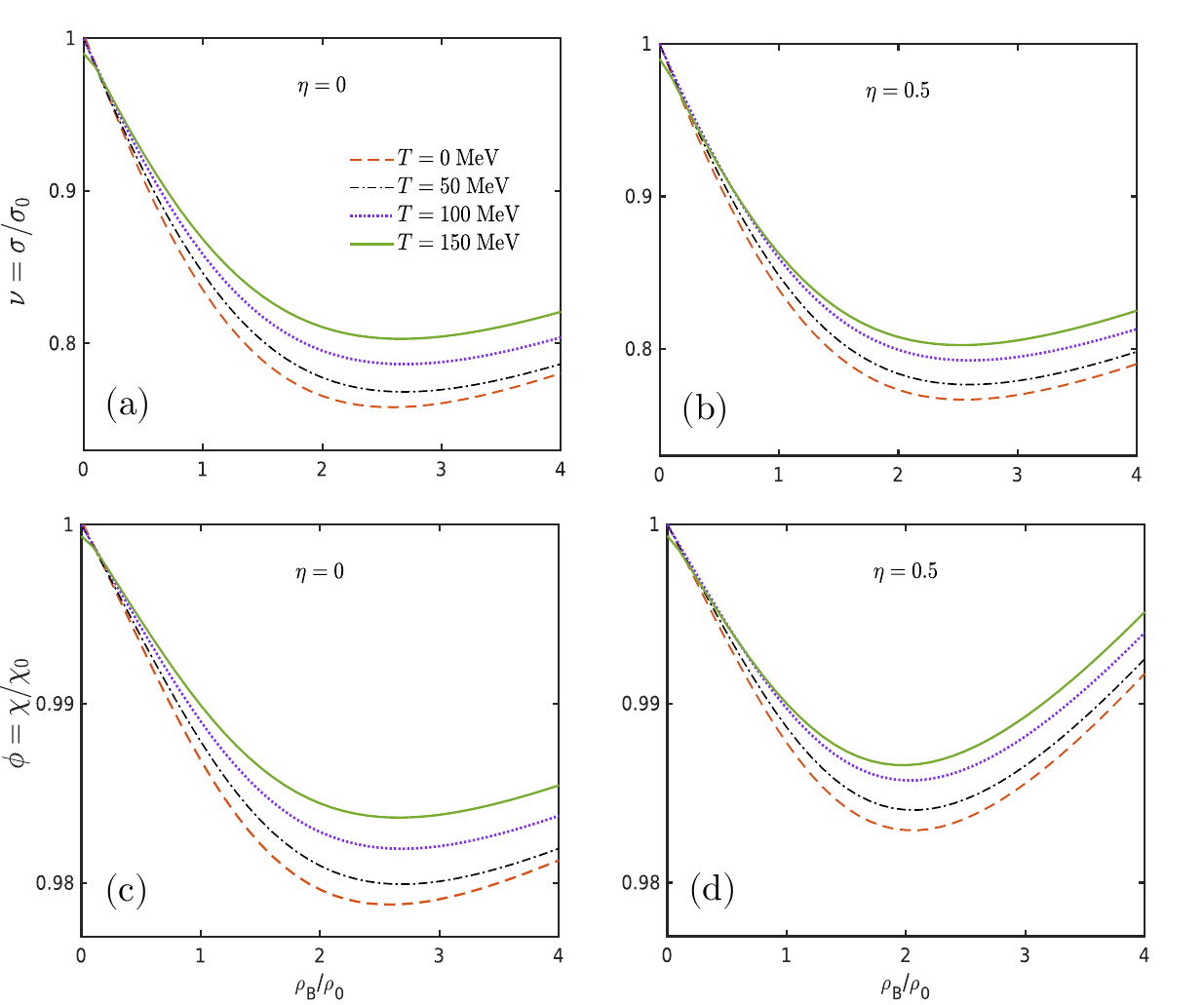}}
\vspace*{8pt}
    \caption{\raggedright{Behavior of the scalar dilaton field $\chi$ and scalar-isoscalar field $\sigma$ in terms of the ratio, $\phi=\chi/\chi_0$ and $\nu =\sigma/\sigma_0$, respectively, in the (a,c) symmetric ($\eta =  0$ ) as well as (b,d) asymmetric ($\eta = 0.5$) nuclear matter, is plotted as a function of the baryon density $\rho_B$ (in units of $\rho_0$) up to $\rho_B=4\rho_0$. We show the results as the temperatures varies from T = 0 MeV to T = 50, 100 and 150 MeV.}}
    \label{fieldNM}
\end{figure}

To begin with, we study the scalar mean field potentials in nuclear matter.
The behaviors of the scalar dilaton field $\chi$ and the scalar isoscalar field $\sigma$ in the different nuclear environments are shown in figs. \ref{fieldNM}(a-d) in terms of the ratio of the fields to its vacuum expectation values ($\chi_0$ or, $\sigma_0$) as $\phi = \chi/\chi_0$ and $\nu = \sigma/\sigma_0$.
Using the parameters mentioned in Table-\ref{table1}, the behaviors of the scalar fields in the nuclear matter are obtained by solving the Eqs.  (\ref{c10})-(\ref{c17}) for hot and dense symmetric (with $\eta=0$) as well as asymmetric (with $\eta ~=~ 0.5$)  nuclear matter.
We restrict ourselves to study up to the temperature T = 150 MeV and density up to $4\rho_0$, with the saturation density, $\rm{\rho_0 = 0.15\; fm^{-3}}$.
In nuclear matter, a nonlinear density profile is observed for the scalar fields, where the value of $\phi$($\nu$) decreases initially. 
After it reaches the minimum, it starts to increase with the density due to the mutual, density-dependent effects of the attractive and repulsive interactions considered in this study. 
Unaltering the overall trend obtained in the dense medium, the values of the fields are increasing as we move towards the higher temperature, illustrated in the figs. \ref{fieldNM}(a-d).
However, it is observed from the fig. \ref{fieldNM} that the amount of the medium modifications to the scalar isoscalar field $\sigma$ (in units of $\sigma_0$) as a function of the nuclear matter density, is larger as compared to the modifications to the scalar dilaton field $\chi$ (in units of $\chi_0$), at zero as well as at finite temperatures.
Furthermore, with the increase of isospin asymmetry in the medium, i.e., when the number densities of the protons and neutrons induce a nonzero isospin density, the in-medium dilaton field shows a much distinct modification compared to the case of mean scalar-isoscalar field in the asymmetric nuclear matter (ANM).
As stated earlier, $\eta$, in the figures, is the isospin asymmetry parameter where $\eta = 0$ represents the symmetric nuclear matter (SNM) and $\eta = 0.5$ depicts the pure neutron matter or the highest degree of asymmetry.
The isospin effects are seen to be large for the higher baryon densities.
On the other hand, the nonzero temperature (T $\neq$ 0) effect is conveyed through the thermal distribution functions, as stated in Eqs. (\ref{c16a})-(\ref{c16b}), in the scalar and the number densities of the nucleons (see Eqs. (\ref{c14})-(\ref{c15})), leading to a modification to the scalar fields.
At finite temperature and zero baryon density (i.e., $\mu_i^*=0$), the Fermi-Dirac distribution functions lead to the nonzero scalar densities, which causes a slight decrease in the scalar fields' ($\sigma$ and $\chi$) magnitudes from their vacuum values.
The effect becomes prominent at higher temperatures, notably after it reaches the critical temperature.
In figs. \ref{fieldNM}(a-d), which is observed at T = 150 MeV where the field ratios attain slightly lesser values than 1 at $\rho_B=0$. 
At finite densities (i.e., $\mu_i^*\neq0$), the combined contribution of higher momenta and finite temperature affects the scalar as well as the number densities.
Therefore, the values of the scalar fields at finite temperatures are larger than their values obtained at zero temperature and increase with the increasing temperature at any fixed baryon density. 
It can be interpreted from the behavior of the fields that their in-medium values remain less than their vacuum expectation values, till $\rho_B=4\rho_0$, indicating a negative mass shift of the heavy quarkonium states in a hot, dense, and asymmetric nuclear matter (see Eq.(\ref{Q4})). 
\subsection{Mass Shifts of the quarkonia states in nuclear matter and finite nuclei} 
\label{B}
\begin{figure}[th]
\centerline{\includegraphics[width=\textwidth]{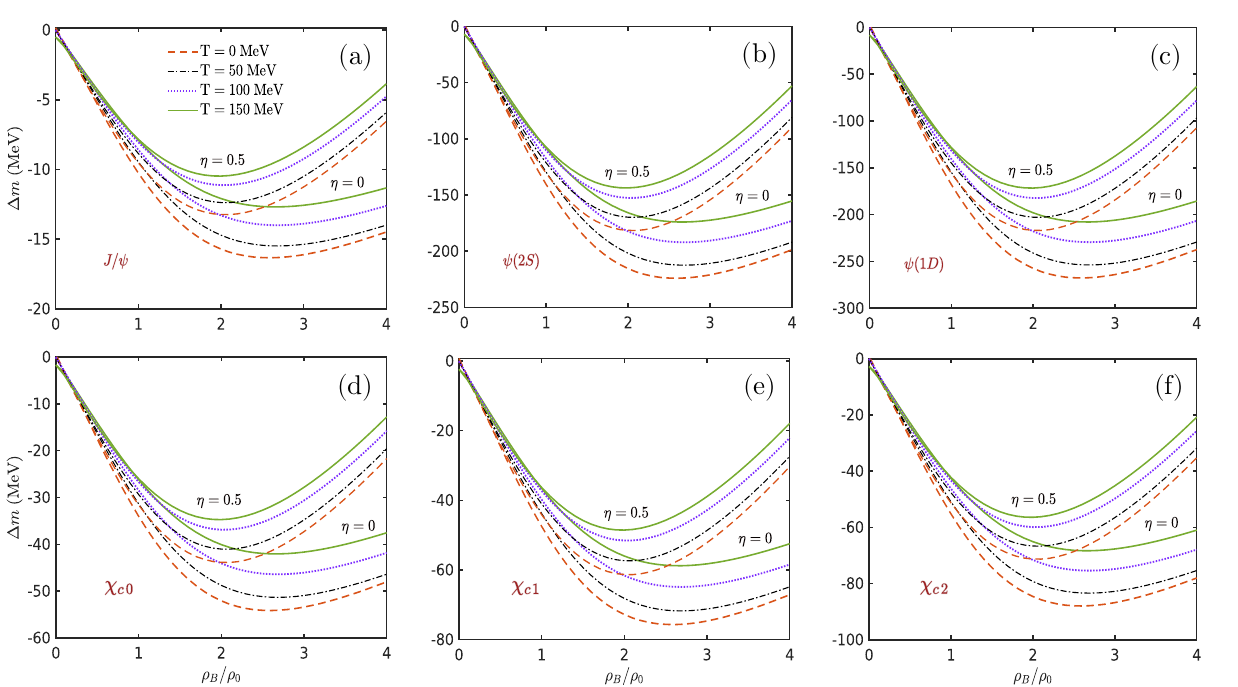}}
\vspace*{8pt}
    \caption{\raggedright{The mass shifts (in MeV) of charmonia states (a) $J/\psi$, (b) $\psi(2S)$, (c) $\psi(1D)$, (d) $\chi_{c0}$, (e) $\chi_{c1}$, and (f) $\chi_{c2}$ in the dense nuclear matter varying from $0$ to $4\rho_0$, are plotted as a function of baryon density $\rho_B$ (in units of $\rho_0$) for the temperatures, T = 0, 50, 100 and 150 MeV. Different conditions of isospin asymmetry, where $\eta=0$ and $\eta=0.5$ represent symmetric and asymmetric nuclear matter, respectively are indicated as lower and upper panels in the plots. }}
    \label{CMassNm}
\end{figure}

\begin{figure}[th]
\centerline{\includegraphics[width=\textwidth]{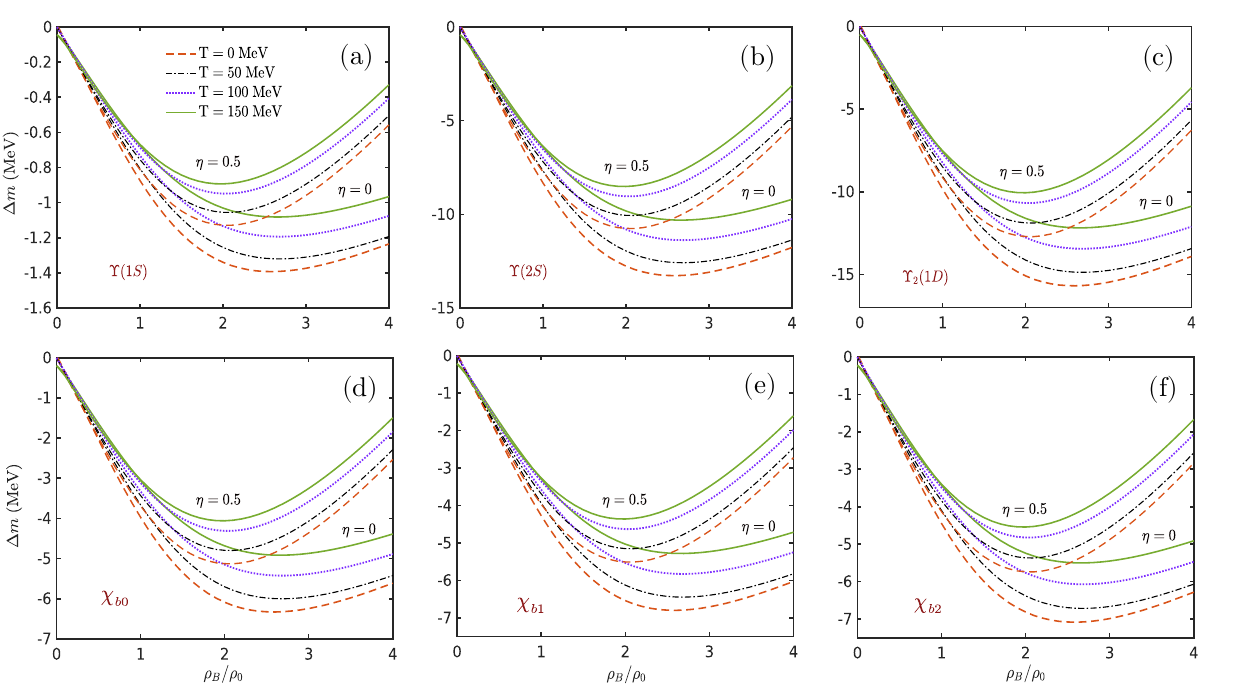}}
\vspace*{8pt}
    \caption{\raggedright{ Same as fig.\ref{CMassNm}, for bottomonia states (a) $\Upsilon(1S)$ (b) $\Upsilon(2S)$ (c) $\Upsilon_2(1D)$ (d) $\chi_{b0}$ (e) $\chi_{b1}$ and (f) $\chi_{b2}$.}}
    \label{BMassNm}
\end{figure}

In this subsection, we present the effects of different nuclear environments on the masses of the charmonium ($J/\psi$, $\psi(2S)$, $\psi(1D)$, $\chi_{c0}$, $\chi_{c1}$ and $\chi_{c2}$) and bottomonium ($\Upsilon(1S)$, $\Upsilon(2S)$, $\Upsilon_2(1D)$, $\chi_{b0}$, $\chi_{b1}$ and $\chi_{b2}$) states within the current framework of the generalized linear sigma model.
As we have discussed, the mass shifts of the quarkonium states are proportional to the difference between the in-medium scalar gluon condensate and its vacuum value with the proportionality constant depending on the binding energy and the dipole size of the corresponding state. 
The scalar gluon condensate is expressed in terms of the fourth power of the scalar dilaton field $\chi$ (see Eq. (\ref{c8})).
By incorporating the obtained in-medium values of the dilaton fields in Eq. (\ref{Q4}), the mass shifts of the quarkonia within the medium can be estimated.
The obtained behaviors within such nuclear environments are illustrated in figs. \ref{CMassNm}-\ref{BMassNm}.
The figures indicate that the variations of the in-medium mass shifts are largely dictated by the dilaton field behavior.
As we can infer from the figs. \ref{fieldNM}(c-d) that within the nuclear matter the in-medium values of dilaton field is lesser than its vacuum values, i.e., $\chi<\chi_0$, and therefore, the masses of the quarkonia states receive a negative shift.
Initially, the masses experience a large amount of mass drop up to around $3\rho_0$ ($2.5\rho_0$) for SNM(ANM), the amount then decreases till $4\rho_0$, irrespective of temperatures.
Due to the relatively higher values of the dilaton field $\chi$ in ANM compared to SNM, the magnitude of the mass shifts decreases in the ANM while moving from SNM.
At the nuclear matter saturation density $\rho_B$ = $\rho_0$, the mass shift of $J/\psi$ state is around -10.29 MeV for $\eta=0$, and -9.52 MeV for $\eta=0.5$.
As indicated in ref. \cite{luke, DAM1}, due to the small binding energies, bottomonium states attain much smaller mass modifications than the charmonium states.
In SNM, at $\rho_0$, the mass drop of $\Upsilon(1S)$ state is almost -0.88 MeV, and at $4\rho_0$, it is -1.23 MeV.
For ANM, it modifies to nearly -0.81 MeV ($\rho_0$) and -0.56 MeV ($4\rho_0$).
In the present investigation, the higher-lying states are observed to have a much larger mass drop as compared to the lower states because of the respective binding energies and the dipole sizes of the states.
The mass shifts (in MeV) of $\psi(2S)$ and $\psi(1D)$ are obtained as -141.12 and -168.59 for $\eta=0$, -130.62 and -156.06 for $\eta=0.5$, at $\rho_B=\rho_0$.
This investigation shows a large mass drop of quarkonium states, compared to the existing studies mentioned in section- \ref{intro}.
However, as the temperature increases, the states experience smaller mass drop due to the combined effects of higher momenta, nonzero effective chemical potentials, and nonzero temperatures.
For ANM these magnitudes are even less than the corresponding $\eta = 0$ results.
At $T\neq0\;\rm{MeV}$, due to the presence of the nontrivial Fermi distribution functions, slight mass shifts of the quarkonia states are obtained at zero density, leading to the hadron masses to be different from their vacuum masses.
The effect becomes prominent at higher temperatures.
For instance, as shown in Figs. \ref{CMassNm} and \ref{BMassNm}, at a temperature of \(T = 150\) MeV, a nonzero mass shift is observed at \(\rho = 0\).

In this study, the values of the strength parameters ($\b$) are determined by the root mean squared radii of the quarkonia states.
For the charmonia states, $J/\psi$,  $\chi_{c1}(1P)$, $\psi(2S)$ and $\psi(1D)$, the rms radii are (0.47 fm)$^2$,(0.74 fm)$^2$, (0.96 fm)$^2$ and (1 fm)$^2$ respectively \cite{eichten_2}, yielding the $\beta$ values as 0.52 GeV, 0.42 GeV, 0.38 GeV and 0.37 GeV \cite{amarvdmesonTprc,amarvepja, F1}. 
Whereas for the bottomonia states, $\Upsilon(1S)$, $\Upsilon(2S)$, $\Upsilon(3S)$ and $\Upsilon(4S)$, the $\b$ values are 1.31 GeV, 0.92 GeV, 0.78 GeV and 0.64 GeV, respectively, obtained from their corresponding rms radii (0.1843 fm)$^2$, (0.4026 fm)$^2$, (0.5925 fm)$^2$ and (0.8449 fm)$^2$ \cite{eichten_2}.
For the remaining states, $\b$ values can be obtained through the linear interpolation (or extrapolation) of the vacuum mass versus $\b$ graph.
The values of $\b$ (in GeV) for $\chi_{c0}$, $\chi_{c2}$, $\chi_{b0}$, $\chi_{b1}$, $\chi_{b2}$ and $\Upsilon_2(1D)$, are determined as 0.45, 0.41, 0.99, 0.97, 0.96, and 0.86, respectively. 
From the values, it can be interpreted that $\b$ values are larger for the smaller size quarkonia states.

\begin{figure}[th]
\centerline{\includegraphics[width=\textwidth]{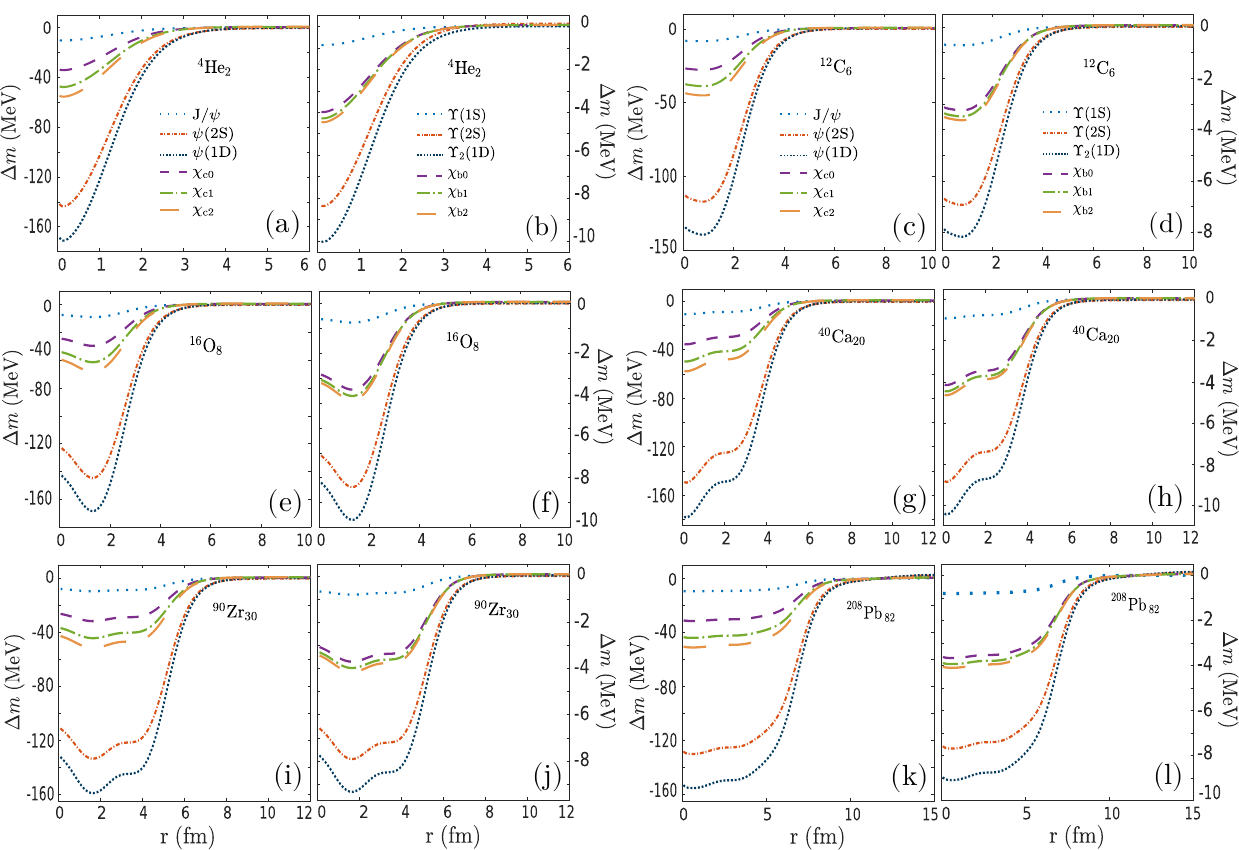}}
\vspace*{8pt}
    \caption{\raggedright{Behavior of the mass shifts (in MeV) of charmonia ($J/\psi$, $\psi(2S)$, $\psi(1D)$, $\chi_{c0}$, $\chi_{c1}$, and $\chi_{c2}$) and bottomonia ($\Upsilon(1S)$, $\Upsilon(2S)$, $\Upsilon_2(1D)$, $\chi_{b0}$, $\chi_{b1}$, and $\chi_{b2}$) states while captured inside (a-b) ${\rm{^{4}He}}$, (c-d) ${\rm{^{12}C}}$, (e-f) ${\rm{^{16}O}}$, (g-h) ${\rm{^{40}Ca}}$, (i-j) ${\rm{^{90}Zr}}$ and (k-l) ${\rm{^{208}Pb}}$ nuclei. }}
    \label{MassFN}
\end{figure}

Once we obtain the behavior of the quarkonia masses in the infinite nuclear matter, using the local density approximation (LDA), the behaviors of the mass shifts within a finite nucleus are estimated.
In figs. \ref{MassFN}(a-l), we present the mass shifts of the different quarkonia states within ${\rm{^{4}He}}$, ${\rm{^{12}C}}$, ${\rm{^{16}O}}$, ${\rm{^{40}Ca}}$, ${\rm{^{90}Zr}}$, and ${\rm{^{208}Pb}}$ nuclei.
The behavior is largely governed by the corresponding nuclear density profiles.
Similar to the results obtained in the nuclear matter study, the magnitudes of $\Delta m$ tend to increase for higher mass hadrons, and at $r \rightarrow 0$, $\Delta m$ represents the values at nuclear saturation density since towards the center, the average baryon density of the nucleus converges to the nuclear matter saturation density ($\rho_0\sim0.15 \;{\rm{fm^{-3}}}$) \cite{me_fn_24}.
At $r\rightarrow0$, the magnitudes of the mass shifts ($\Delta m$) are not significantly influenced by the choice of nuclei.

\subsection{Quarkonia-nuclear bound states}
\label{bound}
\begin{figure}[th]
\centerline{\includegraphics[width=\textwidth]{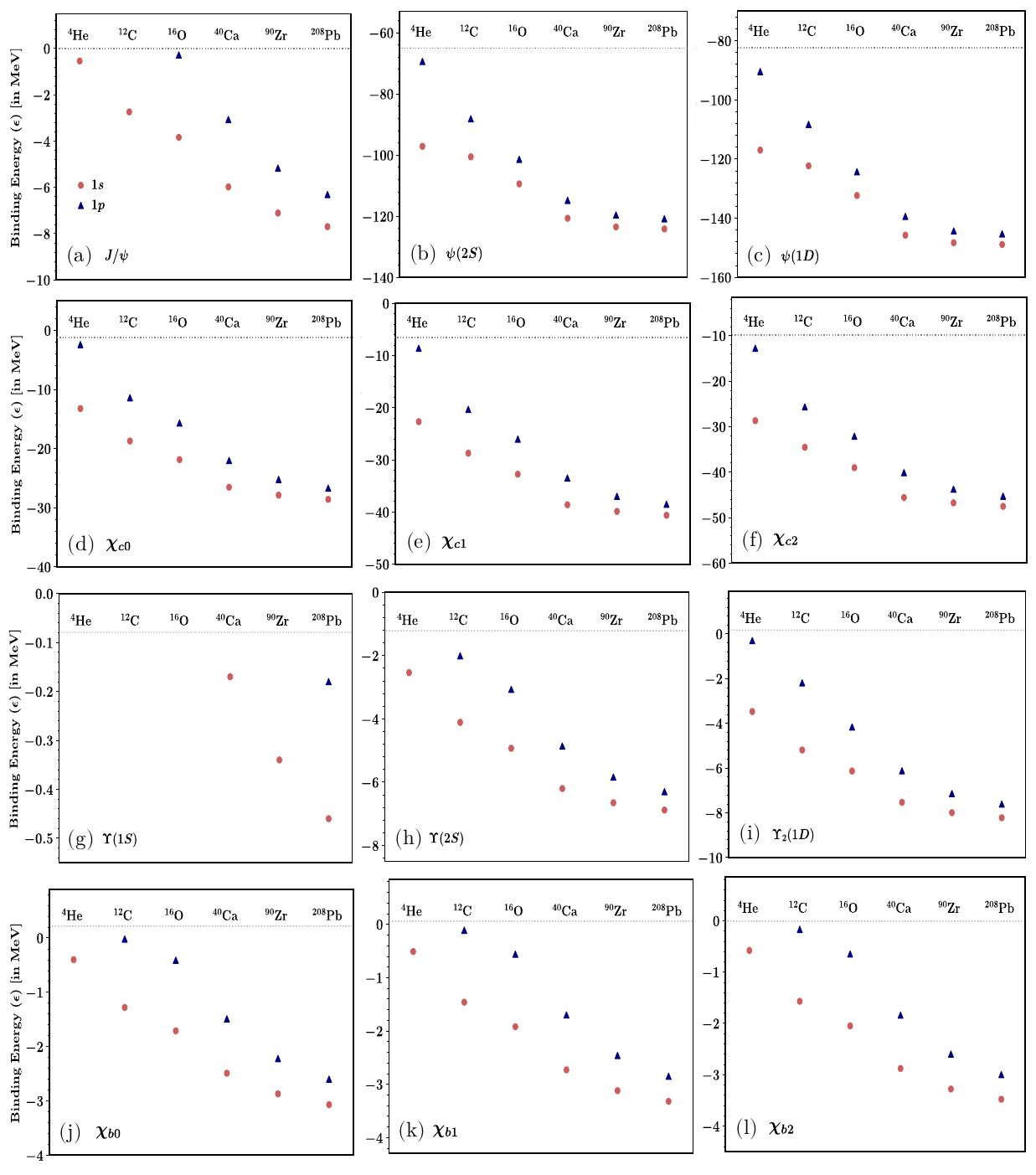}}
\vspace*{8pt}
    \caption{\raggedright{Binding energies (in MeV) of different quarkonia states, (a) $J/\psi$, (b) $\psi(2S)$, (c) $\psi(1D)$, (d) $\chi_{c0}$, (e) $\chi_{c1}$, (f) $\chi_{c2}$, (g) $\Upsilon(1S)$, (h) $\Upsilon(2S)$, (i) $\Upsilon_2(1D)$, (j) $\chi_{b0}$, (k) $\chi_{b1}$, and (l) $\chi_{b2}$
    are predicted by the generalized linear sigma model, for the ${\rm{^{4}He}}$, ${\rm{^{12}C}}$, ${\rm{^{16}O}}$, ${\rm{^{40}Ca}}$, ${\rm{^{90}Zr}}$, and ${\rm{^{208}Pb}}$
    nuclei, as indicated by each column.}}
    \label{BE}
\end{figure}

We finally explore the potential formation of quarkonia bound states with the ${\rm{^{4}He}}$, ${\rm{^{12}C}}$, ${\rm{^{16}O}}$, ${\rm{^{40}Ca}}$, ${\rm{^{90}Zr}}$, and ${\rm{^{208}Pb}}$ nuclei, using the potentials calculated within the generalized linear sigma model.
As discussed in section \ref{qn_pot}, these mesic-nuclei potentials for quarkonia are solely contributed by their medium-modified masses.
The nuclear matter study indicates that the nuclear environment is attractive to these mesons.
This attractive potential may lead to the existence of the bound states when the mesons are captured inside the nucleus. 
In this study, we look for quarkonia-nuclei bound states by solving the Klein-Gordon Eq. (\ref{final}) under the condition that the mesons have zero widths in both free space and within the nucleus, which is a wise choice for heavy quarkonia.
The estimated meson-nucleus bound-state energies for different nuclei are presented in fig. \ref{BE}.
These results indicate that both charmonium and bottomonium states are likely to form bound states with all the nuclei considered.
For instance, for $J/\psi$, bound states are observed only for the 1s state in lighter nuclei, as shown in Fig. \ref{BE}(a).
Whereas, due to the smaller mass shift, $\Upsilon(1S)$ does not form bound states with lighter nuclei, as shown in fig. \ref{BE}(g).
Even in the ${\rm{^{40}Ca}}$ and ${\rm{^{90}Zr}}$ nuclei, only the 1s state is observed.
As the mass number increases, both the magnitude of the binding energies and the number of bound states rises.
Consequently, in the heavier nuclei, these states exhibit more bound states with stronger binding, leading to a more intricate spectrum anticipated for these nuclei.
For each nucleus, we only list the ground states (i.e., 1s and 1p), as the higher states require consideration of spin-orbit interactions, which are not considered in the current study.
Since the bound state energies depend solely on the mass shifts of the quarkonia states within the nuclei, the higher orbital states acquire greater binding energies.
The figures also indicate that the binding energies for charmonium states are greater than those for bottomonium states within the same nuclei, a difference attributed to their respective mass behavior inside the nucleus (see section \ref{B}).
Compared to the findings in ref. \cite{qmc_C,qmc_B}, where the binding energies for quarkonia are strongly dependent on the cut-off momentum, and the quarkonia-nuclei potentials are derived from the 1-loop contributions of different hadronic meson channels within the QMC model, the bound state energies estimated in the current study are lower.
\section{SUMMARY}
\label{summary}
To summarize, we have investigated how the masses of the various charmonia ($J/\psi$, $\psi(2S)$, $\psi(1D)$, $\chi_{c0}$, $\chi_{c1}$, and $\chi_{c2}$) and bottomonia ($\Upsilon(1S)$, $\Upsilon(2S)$, $\Upsilon_2(1D)$, $\chi_{b0}$, $\chi_{b1}$, and $\chi_{b2}$) states are modified within the symmetric as well as asymmetric nuclear matter with different temperatures varying from 0 MeV to 150 MeV.
Due to the absence of light (anti)quark constituents, the masses receive medium modifications from the in-medium gluon condensate, simulated through the scalar dilaton field within the generalized linear sigma model. 
The model incorporates the QCD scale invariance breaking through an effective potential involving logarithmic terms in the scalar dilaton field $\chi$, which generates the scalar gluon condensate by the trace matching condition.
The estimated behaviors of in-medium masses are strongly dictated by the in-medium behavior of the scalar dilaton field $\chi$, calculated by solving the coupled equations of motion obtained from the considered Lagrangian density.
Therefore, similar to the behavior of the $\chi$ field, the estimated in-medium mass drops of the quarkonium states show a non-monotonous decreasing pattern, where the mass drops initially increase with the density up to around $3\rho_0$ ($2.5\rho_0$) for $\eta = 0$ ($0.5$) and later decrease slowly towards the higher-density region.
Compared to SNM ($\eta = 0$), in ANM ($\eta=0.5$), the overall behavior becomes steeper towards the higher density, leading to smaller mass drops.
However, such characteristics are independent of temperatures.
At the same baryon density, the mass drop decreases with increasing temperature from T=0 to 150 MeV, within the medium.
Overall the resulting mass drops for the charmonium states are much greater than the bottomonium states.
Moreover, the mass shifts of the higher orbital states of charmonia and bottomonia at any particular density are larger than the mass shifts of their corresponding lower orbital states, i.e., they are more strongly bound to the nuclear matter.
Given their experimental importance, heavy quarkonia serve as effective gauges of their interactions, acting as valuable probes of the matter produced in the early stages of HIC experiments, where extremely high temperatures and significant asymmetry are anticipated during collisions.
Therefore, studying in-medium heavy quarkonia masses within such nuclear matter will be an important probe for the strongly interacting matter and its complex dynamics.
Notably, since the in-medium mass cannot be observed directly, such in-medium effects can be experimentally examined by analyzing their spectral characteristics.

In the present study, we have explored another important probe for strongly interacting matter, i.e., mesic-nuclei bound states, which can form when the mesons are produced inside a nucleus under (nearly) recoilless kinematics.
A precise measurement of the binding energies of these states would significantly insights into the strongly interacting systems.
The stronger the bound states, the closer the meson stays at the nuclei, letting these states probe even smaller variations in the nuclear matter properties.
The probable formation of the mesic-nucleus bound states of different quarkonia within ${\rm{^{4}He}}$, ${\rm{^{12}C}}$, ${\rm{^{16}O}}$, ${\rm{^{40}Ca}}$, ${\rm{^{90}Zr}}$, and ${\rm{^{208}Pb}}$ nuclei are investigated by solving Klein-Gordon equation for the corresponding quarkonia-nuclei potentials.
The potentials are calculated within the generalized linear sigma model, which has been extended to study the properties within finite nuclei by accounting for the spatial dependencies of the mean fields.
Our findings present the existence of quarkonia-nuclei bound states if the meson-nucleus potential is substantially attractive.
Our study suggests that the charmonia are more likely to form strong bound states with heavier nuclei than the bottomonia since the formation of the bound states is governed only by their modified masses within the nuclear environments, provided by the nuclei.
Furthermore, due to the larger mass drops, the higher-lying states tend to result in more deeply bound states with the nuclei.
Current experimental programs at J-PARC-E29, $\rm{\bar{P}ANDA}@$FAIR, and JLab have included the investigations on bound states, where a slowly moving meson is trapped at the very central region of the nuclei.
These investigations offer a more promising probe to explore subtle nuclear medium effects and provide valuable insights to explore strongly interacting matter created during HIC experiments.


\end{document}